\documentclass[10pt,conference]{IEEEtran}
\usepackage{epsfig,rotating,setspace,latexsym,amsmath,epsf,amssymb,amsfonts,bm,theorem,cite,authblk, bbm,color, multirow, caption, subcaption, booktabs, graphicx, xcolor}
\IEEEoverridecommandlockouts
\allowdisplaybreaks

\newcommand\rev[1]{\textcolor{black}{#1}}

\newcommand{\Real}{\operatorname{Re}}
\newcommand{\Imag}{\operatorname{Im}}

\newcommand{\NAE}{N_{\mathrm{AE}}}     
\newcommand{\Nmax}{N_{\max}}          
\newcommand{\Reff}{R_{\mathrm{eff}}}

\title{Taking Autoencoder Communications to Link Layer: AI-Native Incremental Redundancy for Low Latency}

\author{Yalin E. Sagduyu and Tugba Erpek}
\affil{\normalsize Nexcepta, Gaithersburg, MD, USA}

\date{}
\IEEEoverridecommandlockouts

\begin{document}
\maketitle
\thispagestyle{empty}

\begin{abstract}
The AI-native framework of autoencoder (AE) communications learns an end-to-end PHY-layer mapping from information bits to complex modulated symbols and their recovery from received samples, jointly integrating modulation and coding and potentially outperforming conventional schemes in short-block transmission or under training-test channel mismatch. However, an AE with a fixed blocklength cannot satisfy link-layer latency and reliability requirements when decoding fails. This paper extends AE-based communications to the link layer through a cross-layer design that maintains consistent transmit power normalization under variable-length transmissions, equips the decoder with a feedback signal for ACK/NACK decisions, and orchestrates feedback-driven retransmissions to maximize goodput. We introduce an incremental redundancy method driven by a learned prefix decoder and compare it with a single-shot fixed-latency baseline and a Chase-style repetition baseline. The proposed method introduces stage-wise transmitter branches that explicitly generate redundancy increments while keeping the receiver and feedback unchanged, enabling round-specific redundancy without modifying the receiver. Using a universal training curriculum and a multi-round objective that balances early- and final-round decoding performance, we evaluate goodput across channel and antenna configurations. Results show that the proposed stage-wise incremental method outperforms repetition and single-shot baselines, with gains most pronounced under fading and increasing with the number of antennas.
\end{abstract}


\section{Introduction}
\emph{Autoencoders} (AEs) provide an \emph{AI-native} end-to-end learning framework for PHY-layer communications in which encoder and decoder blocks based on \emph{deep neural networks} (DNNs) are jointly optimized over the wireless channel \cite{OShea2017TCCN,Dorner2018JSTSP, erpek2019deep}. The encoder maps information bits to complex baseband symbols, while the decoder reconstructs the bits from the received samples, enabling joint learning of modulation and channel coding. AEs have been shown to achieve lower error than conventional modulation and coding schemes with model-based detection and decoding, particularly in short-block transmission or under channel-model mismatch
\cite{ait2022waveform, davaslioglu2024end, dorner2022learning, marasinghe2024waveform, bo2024joint}.

Existing AE communication studies largely focus on \emph{PHY-layer designs} that produce fixed-length codewords. When a block is decoded incorrectly, higher layers trigger recovery. In conventional communications, this is handled by automatic repeat request (ARQ) and hybrid ARQ (HARQ). Chase combining accumulates energy by repeating the same coded symbols \cite{Chase1985}, while incremental redundancy (IR) injects fresh parity to improve the decoder state \cite{caire2002throughput}. The design of these \emph{link-layer} mechanisms is tightly coupled to reliability and latency targets, especially in short-packet and ultra-reliable low-latency communication (URLLC) regimes \cite{Durisi2016,nadeem2025low, feng2023evaluating}. ARQ/HARQ designs also extend to multi-antenna systems with different diversity-multiplexing-delay tradeoffs \cite{sidiropoulos2025interference}.

These considerations motivate taking AE-based communications beyond fixed-length PHY output design toward an \emph{adaptive design} with explicit latency objectives at the \emph{link layer}. Such \emph{cross-layer adaptation} is a key capability of AI-native communications, enabling joint PHY and link-layer optimization under dynamic channel and spectrum conditions. Rather than optimizing only block error rate (BLER) at a preset blocklength, there is a need for low-latency AE protocols that operate jointly across PHY and link layers. AE structures have been adapted to task-oriented communications under latency considerations \cite{Sagduyu2024LowLatency}. In this paper, we take AE-based communications to the link layer and study how to structure feedback, retransmissions, and redundancy for \emph{low-latency} operation when the PHY is based on a learned AE. These capabilities are particularly relevant to mission-critical communications, where AI-native AE-based transceivers can adapt to rapidly changing spectrum conditions while meeting stringent latency and reliability requirements.

Doing so raises several challenges: (i) AE transmit power must remain normalized as channel uses vary; otherwise, apparent gains may result from inconsistent energy accounting. (ii) The decoder must provide a calibrated stopping signal for ACK/NACK decisions without requiring prior knowledge of correctness or operating-point-specific tuning. (iii) Under MIMO channels and short packets, the redundancy transmitted after a NACK determines whether retransmissions contribute new information or merely additional energy. (iv) The design should remain universal across channels and operating conditions without retraining the PHY. (v) Evaluation should isolate gains from link-layer control rather than retraining the underlying AE for each protocol.

\begin{figure}[t!]
  \centering
  \includegraphics[width=1.0\columnwidth]{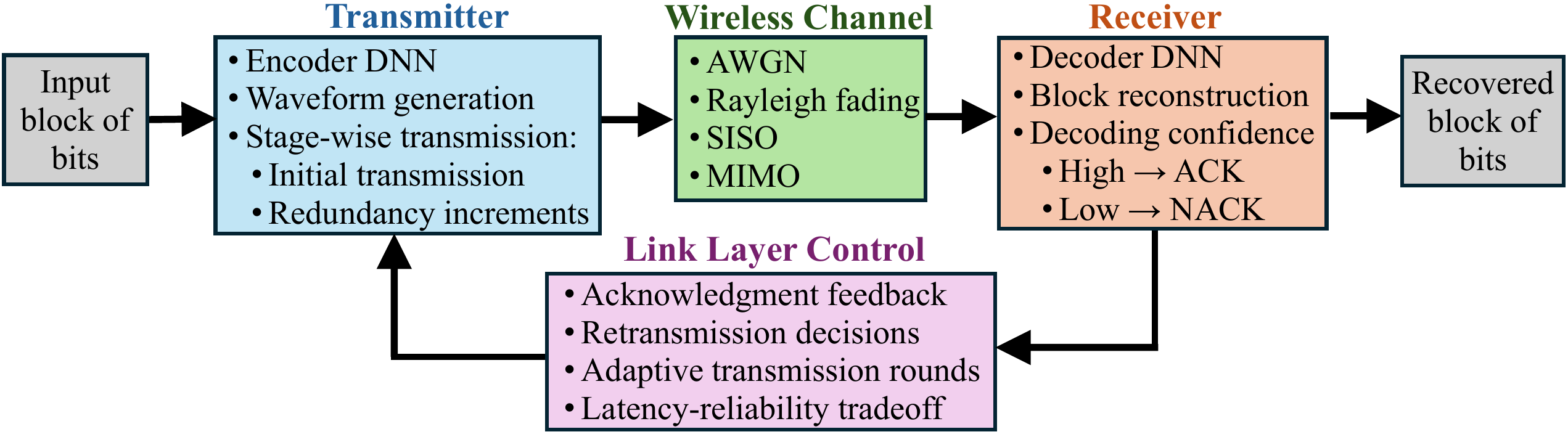}
  \caption{AE communications with PHY- and link-layer integration.}
  \label{fig:system}
\end{figure}

To address these challenges, the AE is embedded in a \emph{cross-layer design} loop (shown in Fig.~\ref{fig:system}) with calibrated stop decisions, an SNR-consistent noise model applied consistently across variable-length transmissions, and \emph{incremental redundancy} delivered in fixed-size rounds to minimize channel uses. We train a single universal AE and evaluate multiple cross-layer control strategies on top of it in order to isolate gains due to link-layer control rather than PHY retraining. 

We develop a cross-layer framework that combines an end-to-end AE PHY with lightweight link-layer control aimed at low latency in addition to high reliability:
(i) we formalize a goodput-centric model that preserves codeword-level energy normalization and \rev{uses a noise model with SNR defined per channel use}, applied consistently across variable-length transmissions;
(ii) we present two minimal-feedback baselines (single-shot fixed-latency and Chase-style repetition) and an incremental redundancy scheme that injects fresh symbols in fixed-size rounds and uses stage-wise transmitter branches to explicitly generate redundancy increments while maintaining the same binary feedback and receiver-side stopping logic;
(iii) we provide a decoder architecture with an explicit stop-branch producing a calibrated confidence for ACK/NACK; and
(iv) we evaluate goodput vs. SNR across different AWGN, Rayleigh fading, SISO and MIMO configurations, and show consistent gains of the proposed method over baselines.

The remainder of the paper is organized as follows. Sec.~\ref{sec:AE} describes AE communications. Sec.~\ref{sec:low_latency} presents low-latency schemes for AE communications. Sec.~\ref{sec:perf} presents performance evaluation results. Sec.~\ref{sec:conclusion} concludes the paper.

\section{Autoencoder Communications} \label{sec:AE}
We consider an end-to-end AE trained universally across a wide SNR range under a mixture of AWGN and flat Rayleigh fading. The transmitter maps a short block of information bits into a length-limited complex sequence on $N_t$ antennas, and the receiver reconstructs the bits after a stochastic channel layer. For stable training and fair comparison across antenna configurations, codeword-level normalization fixes total transmit energy per codeword, and SNR is  mapped to noise variance under variable-length transmissions.

\noindent \textbf{SNR convention.}
SNR is defined per complex channel use under the stated energy normalization. For a transmitted complex symbol with unit average energy per use (after normalization), AWGN is modeled as $\mathcal{CN}(0,\sigma^2)$ with $\sigma^2=(2\cdot 10^{\mathrm{SNR}/10})^{-1}$. Under variable-length operation, the number of channel uses changes the total energy expended, while the per-use SNR definition and noise variance remain unchanged.

\noindent \textbf{Transmission length conventions.}
To avoid symbol conflicts across PHY and link-layer operation, we distinguish the AE waveform horizon, namely the maximum waveform length $\NAE$ (channel uses produced by the learned transmitter for a session), from the total session budget $\Nmax$ (channel uses including possible retransmissions). In the numerical results of Sec.~\ref{sec:perf}, we use $\NAE=\Nmax$ so that the learned transmitter waveform horizon coincides with the session budget. The link-layer algorithms in Sec.~\ref{sec:low_latency} are defined for general $\NAE$, $\Nmax$, initial attempt length, and increment schedule.

\noindent \textbf{System model.} A $k$-bit message $\boldsymbol{b}\in\{0,1\}^k$ is mapped by the encoder to a length-$\NAE$ complex codeword across $N_t$ antennas, where $\boldsymbol{x}_t(\boldsymbol{b})\in\mathbb{C}^{N_t}$ denotes the transmit vector at channel use $t$. The transmitted waveform is represented internally in I/Q form and normalized to ensure unit average transmit energy per channel use. We adopt the discrete-time baseband model
\begin{equation}
\boldsymbol{y}_t
=
\boldsymbol{H}\,\boldsymbol{x}_t(\boldsymbol{b})
+
\boldsymbol{n}_t,
\qquad
\boldsymbol{n}_t\sim\mathcal{CN}
\left(
\boldsymbol{0},
\sigma^2 \boldsymbol{I}_{N_r}
\right),
\end{equation}
where $t$ denotes the channel-use index. We use two channel families throughout the paper: (i) AWGN SISO with $\boldsymbol{H}=1$; and (ii) block Rayleigh MIMO with $\boldsymbol{H}\in\mathbb{C}^{N_r\times N_t}$ having i.i.d.\ $\mathcal{CN}(0,1)$ entries, constant within a block and independent across blocks. This convention, combined with codeword-level transmit-energy normalization, yields SNR comparisons across $N_t$ and $N_r$ under a fixed total transmit-energy-per-use convention. Increasing $N_r$ provides additional noisy observations of the same transmitted waveform. Performance improvements with more receive antennas follow from receiver processing of these additional samples rather than from any change in per-antenna noise variance. Stacking the per-use complex receive vectors into a real-valued tensor yields $\boldsymbol{Y}\in\mathbb{R}^{\NAE\times N_r\times 2}$ with entries separated to real and imaginary components
\begin{equation}
\boldsymbol{Y}[t,:,0]=\Real\{\boldsymbol{y}_t\},\qquad \boldsymbol{Y}[t,:,1]=\Imag\{\boldsymbol{y}_t\}.
\end{equation}
For a length-$L$ decoding attempt with $1\le L\le \NAE$, the effective code rate is $\Reff=k/L$, which is used to describe how many information bits are delivered per accumulated channel use under variable-length operation. Taking $\boldsymbol{Y}$ as input, the decoder performs per-codeword root mean square (RMS) normalization and temporal layer normalization, reshapes the received tensor into a concatenated time-antenna sequence with two I/Q channels, which is further processed to produce bit logits $\boldsymbol{\ell}\in\mathbb{R}^k$ and a scalar stop-branch logit $h\in\mathbb{R}$. We define the stop score $s=\sigma(h)\in[0,1]$, where $\sigma(u)=(1+e^{-u})^{-1}$. This score is used to generate the stopping signal. Reconstructed bits are $\hat{b}_i=\mathbf{1}\{\ell_i>0\}$.

\noindent \textbf{Encoder model.}
The transmitter is a temporal generator that maps the bit vector to an $[\NAE, N_t, 2]$ I/Q tensor. Bits are first expanded into a feature map laid across the $\NAE$ uses so the model can learn time structure directly. A sequence of residual one-dimensional convolutional blocks with GELU activations refines these features into the transmitted waveform. The $1\times 1$ temporal projection produces the $2N_t$ output channels interpreted as I/Q per antenna and time. A codeword-level energy normalization fixes the  transmit energy to $\NAE$. Lightweight dropout is applied in residual blocks during training to regularize the encoder, while evaluation runs deterministically.

\noindent \textbf{Decoder model.}
The receiver is an invariance-rich temporal classifier that maps an accumulated received I/Q tensor to $k$ logits. The decoder operates without explicit channel state information. It begins with per-codeword RMS normalization across all channel uses and receive antennas to remove large-scale gain variations, followed by temporal layer normalization along the concatenated time-antenna sequence to stabilize feature scales within the codeword. The normalized I/Q sequence is fed to a temporal stem convolution that lifts it to a higher-width representation. An eight-block residual one-dimensional convolutional stack with dilation then aggregates evidence efficiently across the entire codeword and across antennas, and squeeze-excitation gating adaptively reweights channels using global context. After feature extraction, global average pooling over time yields a compact embedding that a linear branch maps to $k$ logits and the stop-branch $h$. The design scales naturally with $N_r$. As the number of receive antennas increases, the resulting time-antenna sequence becomes longer, while the same decoder processing pipeline is used. 

\noindent \textbf{Training.} The training loss is the expected bitwise binary cross-entropy between the decoder logits $\boldsymbol{\ell}$ and the ground-truth bits $\boldsymbol{b}$, averaged over the joint distribution of channels and SNRs. Inference decodes each bit by thresholding its logit at zero, which allows direct computation of BLER as the fraction of codewords containing any bit error. After training, dropout is disabled, and the same normalizations used during training are preserved. Evaluation sweeps SNR across $[-2,20]$ dB with large bit budgets ($10^8$) to reduce Monte Carlo variance so that differences between channel and antenna configurations are statistically meaningful. The encoder width is $C_{\mathrm{tx}}=64$ with five residual temporal blocks and kernel size seven, while the decoder width is $C_{\mathrm{rx}}=128$ with eight residual temporal blocks of the same kernel size. Encoder and decoder are optimized jointly. The optimizer is AdamW with a base learning rate of $10^{-3}$, momentum parameters $(0.9,0.99)$, and weight decay $10^{-4}$. Mini-batches contain 1024 codewords. 

The AE is trained with PyTorch on NVIDIA RTX PRO 6000 Blackwell Max-Q Workstation GPUs. The number of trainable parameters in the AE system depends on \(N_t\) and \(N_r\), but varies only slightly with the antenna configuration. For \(N_t=N_r=4\), the AE contains approximately \(2.1\) million trainable parameters, including \(0.3\) million in the encoder and \(1.8\) million in the decoder.

Training uses a universal SNR curriculum implemented as a per-sample SNR sampler over the full range $[-2,20]$ dB with a mid-SNR emphasis. For each mini-batch, every sample draws its own SNR. Training batches contain samples with different SNR values, with most drawn from the mid-SNR region where errors provide informative learning signals, while some samples cover low and high SNR values. For this purpose, with probability $p_{\text{mid}}=0.6$ the value is taken from a window centered at $6$ dB whose width is $0.35$ of the total span, and with probability $0.4$ it is drawn uniformly over the entire $[-2,20]$ dB interval. The same curriculum is applied regardless of whether the forward pass uses AWGN or Rayleigh fading. Optimization follows a warmup-cosine schedule with gradient clipping, mixed precision, and batch size 1024. For evaluation, the encoder is treated as defining a length-$\NAE$ code from which prefixes of length $L\le \NAE$ are drawn, with variable-length decoding supported by global pooling.

\begin{figure}[t!]
  \centering
  \includegraphics[width=0.83\columnwidth]{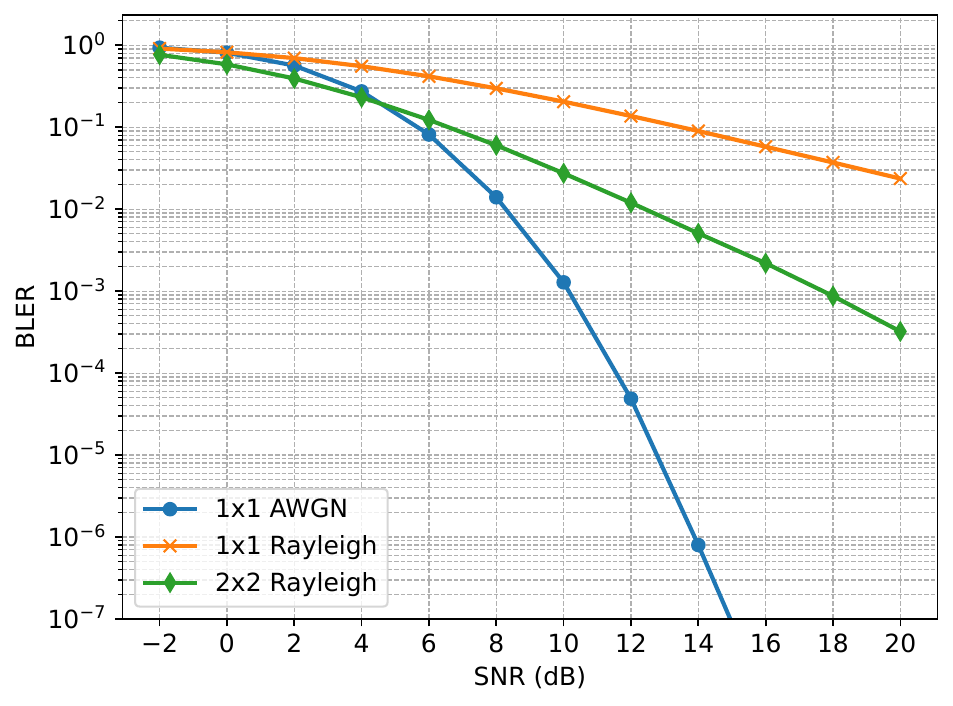}
  \caption{BLER vs.\ SNR for SISO and MIMO communications under AWGN and Rayleigh channels for a single initial AE transmission attempt.}
  \label{fig:AESISOMIMO_BLER}
\end{figure}

\noindent \textbf{Baseline performance.} Fig.~\ref{fig:AESISOMIMO_BLER} shows BLER vs.\ SNR for SISO and MIMO communications under AWGN and Rayleigh channels for a single transmission attempt when $k=16$ bits per block are transmitted once over $\NAE = 16$ channel uses. The BLER drops as SNR, $N_t$, and $N_r$ increase. For reliable communications under strict latency, a retransmission or incremental redundancy is needed when a block results in error. In Sec.~\ref{sec:low_latency}, we investigate low-latency methods for successfully delivering blocks under a session budget $\Nmax\ge \NAE$.

\section{Low-Latency Methods}\label{sec:low_latency}
The receiver returns a one-bit ACK/NACK after each increment with negligible latency and error, as commonly assumed in link-layer analysis. This abstraction isolates payload retransmission, redundancy design, and stopping behavior at the link layer without modeling feedback-transport errors. Power is normalized so that a transmission attempt of length $L$ carries total energy proportional to $L$.

For a given SNR point, let $P_{\mathrm{BLER}}$ denote the block error probability at session termination. Let $U$ denote the number of channel uses consumed in that session, and let $\bar{U}=\mathbb{E}[U]$. Goodput is defined as
\begin{equation}
G \;=\; \frac{k}{\bar{U}} \bigl(1-P_{\mathrm{BLER}}\bigr)\quad \text{bits per channel use}.
\end{equation}

Each block carries $k$ information bits $\boldsymbol{b}\in\{0,1\}^k$. A baseband encoder maps $\boldsymbol{b}$ to a length-$\NAE$ codeword
\begin{equation}
\boldsymbol{x}(\boldsymbol{b}) \;=\; \bigl[\boldsymbol{x}_1(\boldsymbol{b}),\ldots,\boldsymbol{x}_{\NAE}(\boldsymbol{b})\bigr], \qquad \boldsymbol{x}_t(\boldsymbol{b})\in\mathbb{C}^{N_t},
\end{equation}
subject to an energy constraint $\sum_{t=1}^{\NAE} \mathbb{E}\|\boldsymbol{x}_t(\boldsymbol{b})\|_2^2 = \NAE$.
The received sample sequence $\{\boldsymbol{y}_t\}$ and its tensorization $\boldsymbol{Y}$ follow the system model of Sec.~\ref{sec:AE}.

\noindent \textbf{Schedules and antenna placement.}
We describe variable-length transmissions through the set of used time indices $\mathcal{T}\subseteq\{1,\ldots,\Nmax\}$, where $t\in\mathcal{T}$ means the transmitter uses the $t$th channel use (all $N_t$ antennas may be active within the AE waveform at that time). The session latency is measured in channel uses as $U=|\mathcal{T}|$, i.e., the number of channel uses until termination. Energy normalization follows Sec.~\ref{sec:AE}. The transmitter waveform of length $\NAE$ is normalized once at the codeword level so that $\sum_{t=1}^{\NAE}\mathbb{E}\|\boldsymbol{x}_t(\boldsymbol{b})\|_2^2=\NAE$. Variable-length operation uses accumulated prefixes of this globally normalized waveform without additional per-round renormalization. With $L$ accumulated channel uses, the effective code rate is $\Reff=k/L$. 

The receiver uses a decoder
\begin{equation}
(\widehat{\boldsymbol{b}},\, h) \;=\; g\!\left(\{\boldsymbol{y}_t\}_{t\in\mathcal{T}}\right),\qquad s=\sigma(h)\in[0,1],
\end{equation}
that outputs a hard estimate $\widehat{\boldsymbol{b}}$ and a scalar stop-branch logit $h$. A one-bit feedback is generated by the stopping rule
\begin{equation}
\text{ACK} \iff s \ge \tau,\qquad \text{NACK} \iff s<\tau,
\end{equation}
with a fixed threshold $\tau\in(0,1)$. A single $\tau$ is used across all SNRs, channel models, and antenna configurations. The same stop-branch and threshold are used unchanged across all methods to ensure a fair comparison. A prefix of length $L$ denotes the first $L$ channel uses. Transmission proceeds in rounds. An initial prefix of length $L_0$ is followed by increments of size $\Delta$ upon a NACK.

We use an accumulated observation view $\boldsymbol{Y}_{1:L}\triangleq\{\boldsymbol{y}_t\}_{t=1}^{L}$, i.e., the first $L$ channel uses of the received tensor $\boldsymbol{Y}$, where $L$ is the total number of MIMO channel uses so far. In incremental schemes, $L$ grows in rounds. An initial transmission of length $L_0$ is followed by increments of size $\Delta$ until the budget $\Nmax$ is reached. The algorithmic definitions below treat $L_0$, $\Delta$, and $\Nmax$ as free parameters. When an increment size is selectable, $\Delta$ may be chosen from an allowed set (e.g., $\Delta\in\{1,\ldots,L_{\max}\}$) subject to the remaining budget.

We consider two baseline methods and propose a method to improve goodput.

\noindent \textbf{Baseline~1: One-shot fixed-latency transmission.}
Baseline~1 transmits without receiver feedback or adaptivity and terminates after a fixed number of channel uses equal to the session budget $\Nmax$. Since the learned PHY produces a length-$\NAE$ codeword, Baseline~1 uses the first $\Nmax$ channel uses when $\NAE\ge \Nmax$. When $\NAE<\Nmax$, a deterministic padding rule may be used to fill the remaining budget (e.g., repeating the available waveform prefix), but in the numerical results of Sec.~\ref{sec:perf} we use $\NAE=\Nmax$ so no padding is required. The receiver performs a single decoding attempt at the end of the session. This scheme yields deterministic latency equal to $\Nmax$ channel uses and serves as a non-adaptive reference.

Mathematically, Baseline~1 uses a fixed time index set $\mathcal{T}=\{1,\ldots,\Nmax\}$. The receiver forms $\boldsymbol{Y}_{1:\Nmax}$ and computes $(\widehat{\boldsymbol{b}},h)=g(\boldsymbol{Y}_{1:\Nmax})$, with $s=\sigma(h)$. The session stops after this single attempt; thus $U=\Nmax$ and goodput is
\begin{equation}
G_{\text{B1}} = \frac{k\bigl(1-P_{\text{BLER}}^{(\Nmax)}\bigr)}{\Nmax},
\end{equation}
where $P_{\text{BLER}}^{(\Nmax)}$ is the block error probability under the length-$\Nmax$ accumulated observation.

\noindent \textbf{Baseline~2: Conditional repeat with accumulation and re-decoding.}
Baseline~2 introduces minimal feedback and energy accumulation by repeating the same initial burst until the receiver declares success or the budget is exhausted. We apply Chase-style repetition via accumulation of the received observations across repeats and re-decoding after each repetition. The transmitter first sends a length-$L_0$ burst that encodes the entire block, with $1\le L_0\le \NAE$. Upon a NACK, the transmitter retransmits the identical initial prefix of length $L_0$, or the longest feasible portion if fewer than $L_0$ symbols remain in the budget. The receiver combines all repetitions (Chase combining) before re-decoding after each repetition. The procedure continues in steps of $L_0$ channel uses until an ACK is issued or the total transmitted length reaches $\Nmax$. Latency adapts to channel quality through repetition gain, but no new information is introduced across rounds, so diversity beyond SNR accumulation is limited.

Mathematically, we fix an initial burst length $L_0$ and define the first-shot time index set $\mathcal{T}_0=\{1,\ldots,L_0\}$. The transmitter constructs $\{\boldsymbol{x}_t(\boldsymbol{b})\}_{t=1}^{L_0}$ and repeats it upon NACK. After the $j$th repetition, with accumulated length
\begin{equation}
L_j=\min\{jL_0,\Nmax\},
\end{equation}
the receiver fuses the $j$ observations and updates its decision:
\begin{equation}
(\widehat{\boldsymbol{b}}^{(j)}, h^{(j)}) = g\bigl(\boldsymbol{Y}_{1:L_j}^{\text{acc}}\bigr),\qquad s^{(j)}=\sigma\!\bigl(h^{(j)}\bigr).
\end{equation}
Transmission stops at the first $j$ such that $s^{(j)}\ge\tau$ or when $L_j=\Nmax$. Consequently $U=L_J$ for a stopping index $J$. Because the repeated content is identical, the accumulated observation energy increases with $j$ under combining, but no new redundancy is introduced. The induced goodput follows the global definition $G_{\text{B2}} = \frac{k}{\mathbb{E}[U]}(1-P_{\mathrm{BLER}})$, where $P_{\mathrm{BLER}}$ and $\mathbb{E}[U]$ are computed under Baseline~2 operation.

\noindent \textbf{Proposed method: Stage-wise incremental redundancy with per-round transmitter branches.}
The proposed method targets the core inefficiency of repetition-based recovery. Once the receiver has already observed a burst, retransmitting the same waveform predominantly increases energy without necessarily creating new, informationally complementary observations under short-packet decoding. The proposed design injects fresh redundancy in a controlled, round-based way while preserving minimal feedback and a fixed receiver decision rule.

The procedure operates in rounds with a configurable prefix schedule. An initial transmission of length $L_0$ is followed, upon NACK, by one or more increments of length $\Delta$ until either the receiver declares success or the session budget $\Nmax$ is exhausted. After each round, the receiver re-decodes using all accumulated observations $\boldsymbol{Y}_{1:L_r}$ and applies the same stop-branch threshold test $s^{(r)}\ge \tau$. This retains the link-layer interface of one-bit ACK/NACK and does not require any change in receiver architecture, receiver-side inference beyond re-decoding, or feedback format compared to Baseline~2.

The distinguishing feature is the transmitter parameterization. Rather than relying on a single monolithic mapping whose long waveform is later partitioned into increments, the transmitter explicitly generates each round's increment via a stage-specific branch. A shared bit-to-feature expansion first produces a latent temporal representation over the full horizon of $\NAE$ uses, capturing the global structure of the codeword implied by the message bits. Then, the stage-specific convolutional stacks map this shared representation into (i) the initial prefix of length $L_0$ and (ii) each subsequent increment of length $\Delta$. The transmitted waveform is assembled by taking the stage-$0$ output on $[1:L_0]$ and stage-$r$ outputs on $[L_{r-1}+1:L_r]$ for $r\ge 1$, with zeros elsewhere. This design specializes each increment to the receiver state at that round (i.e., conditioned on earlier decoding failures) while keeping receiver-side stopping, combining, and decoding unchanged. The resulting redundancy mechanism remains feedback-efficient, supports goodput maximization under strict latency budgets, and can be integrated with other AE architectures.

Formally, we define round lengths $\mathcal{L}=\{L_0,L_1,\ldots,L_R\}$ with $L_r=L_0+r\Delta$ and $L_R=\Nmax$, where $L_R\le \NAE$ by construction. The proposed transmitter defines a collection of stage mappings $\{\phi_r\}_{r=0}^{R}$ that each produce a length-$\NAE$ waveform $\boldsymbol{x}^{(r)}(\boldsymbol{b})$ over $t\in\{1,\ldots,\NAE\}$, where only the portion intended for stage $r$ is used. The actual transmitted codeword is assembled as
\begin{equation}
\boldsymbol{x}_t(\boldsymbol{b}) \;=\;
\begin{cases}
\boldsymbol{x}^{(0)}_t(\boldsymbol{b}), & 1\le t \le L_0,\\
\boldsymbol{x}^{(r)}_t(\boldsymbol{b}), & L_{r-1}< t \le L_r,\quad r=1,\ldots,R.
\end{cases}
\end{equation}
The link-layer procedure is as follows. At round $r\in\{0,\ldots,R\}$, the transmitter has sent the accumulated prefix $\boldsymbol{x}_{1:L_r}(\boldsymbol{b})$. The receiver computes
\begin{equation}
(\widehat{\boldsymbol{b}}^{(r)}, h^{(r)}) \;=\; g\!\bigl(\boldsymbol{Y}_{1:L_r}\bigr),\qquad s^{(r)}=\sigma(h^{(r)}).
\end{equation}
If $s^{(r)}\ge \tau$, the receiver sends ACK and the session terminates with $U=L_r$. Otherwise, if $r<R$, the receiver sends NACK and the transmitter sends the next increment $\boldsymbol{x}_{L_r+1:L_{r+1}}(\boldsymbol{b})$. The session terminates at the first round $R^\star$ with $s^{(R^\star)}\ge\tau$, or at $R$ when the budget is exhausted. The induced goodput follows the global definition $G_{\text{P}} = \frac{k}{\mathbb{E}[U]}(1-P_{\mathrm{BLER}})$, where $P_{\mathrm{BLER}}$ and $\mathbb{E}[U]$ are computed under the proposed method.

\noindent \textbf{Multi-round training objective for the proposed method.}
The proposed incremental method relies on decoder performance at intermediate round lengths $L_0,L_1,\ldots,L_{R-1}$ as well as at the final length $L_R=\Nmax$. Accordingly, training uses a multi-round loss that combines (i) an average intermediate-round bit loss and (ii) a final-round bit loss, with fixed total weights that do not depend on the number of rounds. Let $\mathcal{L}_{\mathrm{bce}}(\boldsymbol{\ell}^{(r)},\boldsymbol{b})$ be the bitwise binary cross-entropy at round $r$. With $R+1$ rounds and weights $w_L,w_N\ge 0$ (normalized to sum to one), the combined bit loss is
\begin{equation}
\mathcal{L}_{\mathrm{bits}}
\;=\;
w_L \cdot \frac{1}{R}\sum_{r=0}^{R-1}\mathcal{L}_{\mathrm{bce}}(\boldsymbol{\ell}^{(r)},\boldsymbol{b})
\;+\;
w_N \cdot \mathcal{L}_{\mathrm{bce}}(\boldsymbol{\ell}^{(R)},\boldsymbol{b}).
\end{equation}
If the stop-branch is enabled, it is trained to predict whether the current-round hard decision is fully correct, using a binary cross-entropy loss $\mathcal{L}_{\mathrm{stop}}$ and an additional weight $w_{\mathrm{stop}}$:
\begin{equation}
\mathcal{L}\;=\;\mathcal{L}_{\mathrm{bits}} \;+\; w_{\mathrm{stop}}\,\mathcal{L}_{\mathrm{stop}}.
\end{equation}

This objective aligns training with link-layer operation. Intermediate rounds promote early decoding to reduce latency, while the final round preserves reliability.

The overall design balances reliability and efficiency. Baseline~1 sacrifices efficiency to guarantee a fixed latency. Baseline~2 recovers some efficiency by stopping early on easy channel realizations but suffers from diminishing returns because repeats add energy without new redundancy. The proposed method addresses diminishing returns by explicitly generating fresh redundancy in fixed-size increments, and by specializing each increment to its transmission round through stage-wise transmitter branches, while maintaining the same binary feedback, receiver architecture, and stopping rule.

\noindent \textbf{Computational and feedback overhead.}
Baseline~1 requires no feedback and a single decoding per session. Baseline~2 requires one-bit ACK/NACK feedback and up to $\lceil \Nmax/L_0 \rceil$ decoding attempts with combining across repeats. The proposed method requires the same one-bit feedback and up to $R$ intermediate decoding attempts (one per increment); it differs from Baseline~2 primarily in that retransmissions transmit new symbols rather than repeats. Relative to transmitter architectures that produce a single waveform, the proposed stage-wise construction adds modest transmitter-side complexity via stage branches but no added receiver-side inference beyond the decoding already required to produce ACK/NACK.

\section{Performance Evaluation}\label{sec:perf}
\begin{figure}[t!]
  \centering
  \includegraphics[width=0.83\columnwidth]{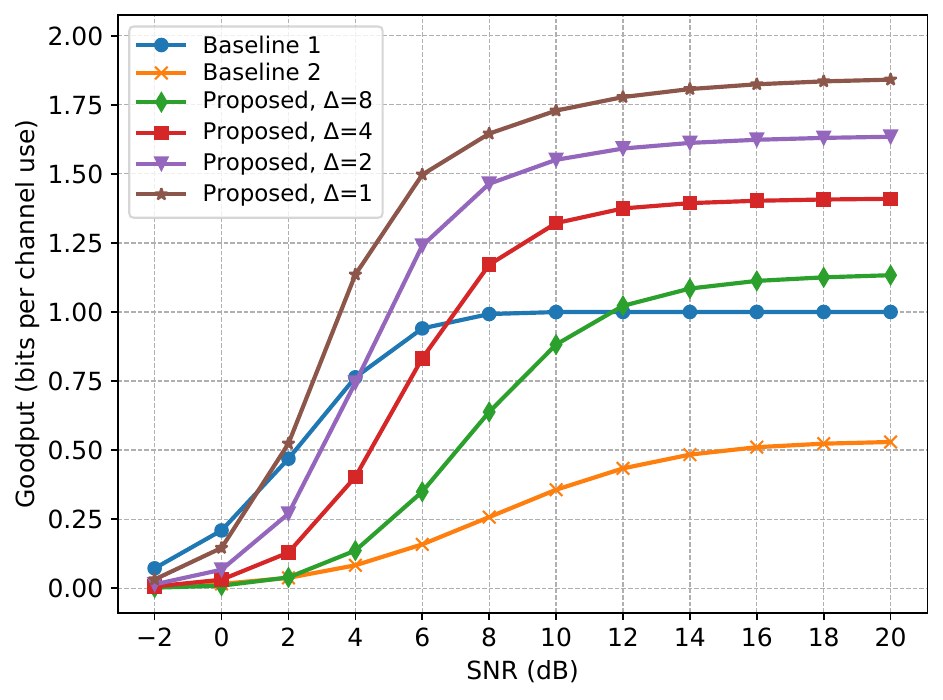}
  \caption{Goodput vs.\ SNR for 1$\times$1 AWGN channel.}
  \label{fig:incremental_1x1AWGN}
  \vspace{0.2cm}

  \centering
  \includegraphics[width=0.83\columnwidth]{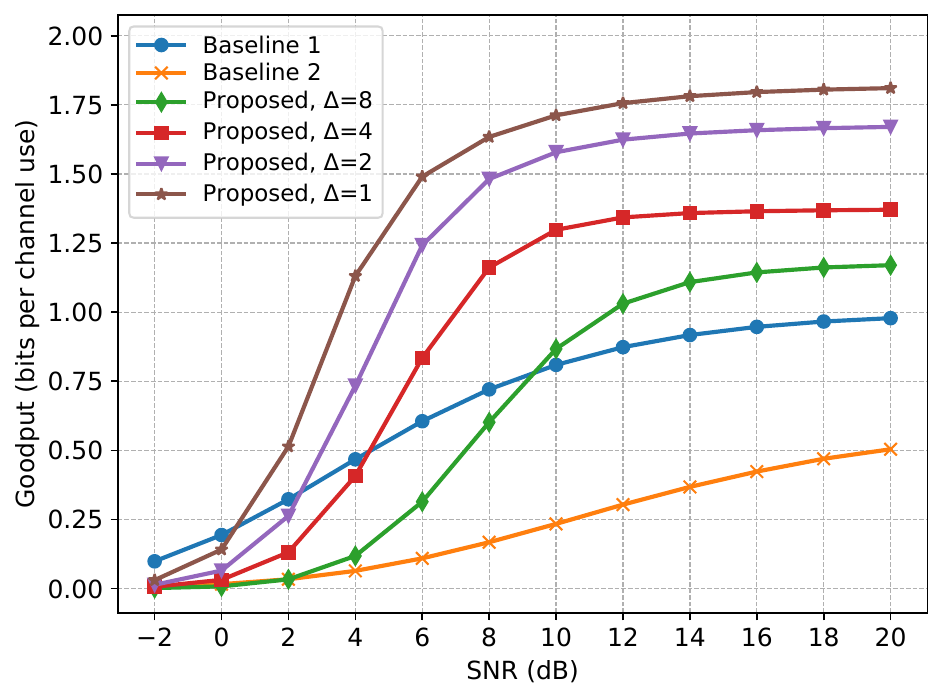}
  \caption{Goodput vs.\ SNR for 1$\times$1 Rayleigh channel.}
  \label{fig:incremental_1x1Rayleigh}
  \vspace{0.2cm}

  \centering
 \includegraphics[width=0.83\columnwidth]{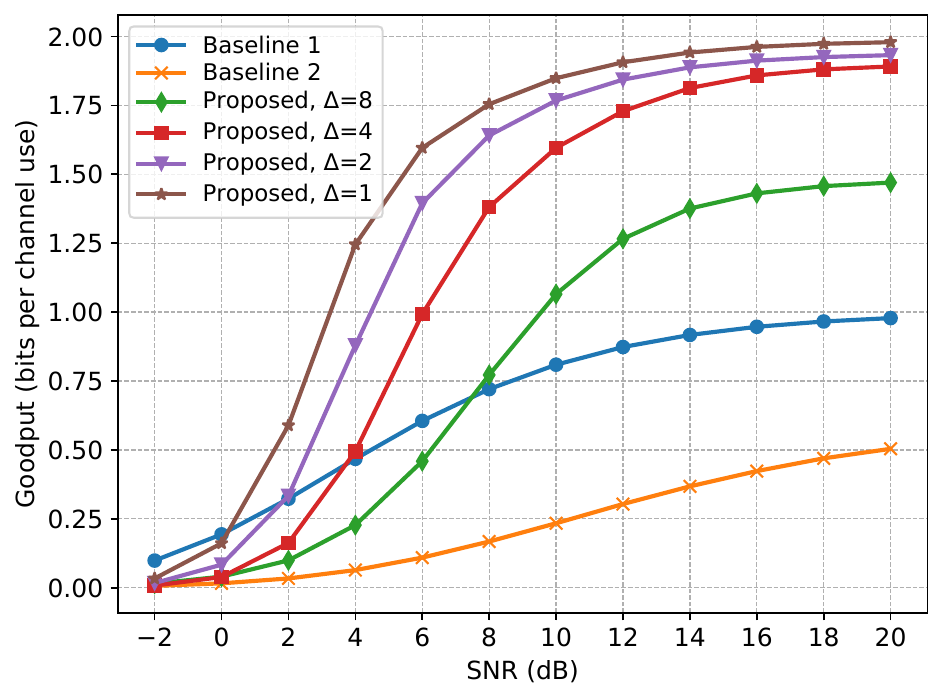}
  \caption{Goodput vs.\ SNR for 2$\times$2 Rayleigh channel.}
  \label{fig:incremental_2x2Rayleigh}
  \vspace{-0.3cm}
\end{figure}

Different methods are compared at common SNR points with identical power normalization. For each point, $10^8$ blocks are transmitted to estimate BLER, average channel uses, and goodput. The channel model is held fixed within a block for block-fading scenarios and regenerated independently across blocks. Fig.~\ref{fig:incremental_1x1AWGN}-\ref{fig:incremental_2x2Rayleigh} show goodput vs.\ SNR for 1$\times$1 SISO AWGN, 1$\times$1 SISO Rayleigh, and 2$\times$2 MIMO Rayleigh configurations. Results are averaged over seeds to smooth variability across repetitions. $k=16$ bits are transmitted per block. Baseline~2 and the proposed method use an initial transmission length $L_0=8$ channel uses, while the total session budget is $\Nmax=16$ channel uses. The proposed method uses increments chosen from $\Delta\in\{1,2,4,8\}$, subject to the remaining budget. Larger $\Delta$ reduces decoding and feedback rounds at the cost of coarser stopping granularity. In these results, we use $\NAE=\Nmax=16$ so that the learned transmitter waveform horizon coincides with the session budget.

For each fixed channel type, antenna configuration, and SNR, the goodput ranking is consistent, with the proposed method achieving the highest goodput, followed by Baseline~2, and finally Baseline~1. For the SISO 1$\times$1 case, goodput drops from AWGN to Rayleigh fading channel models. Under Rayleigh channels, increasing the number of transmit and receive antenna elements yields substantial goodput gains due to increased spatial diversity. Overall, Baseline~1 provides a non-adaptive fixed-latency reference. Baseline~2 introduces feedback-driven repetition that adapts latency through combining but adds no new information. The proposed method strengthens incremental redundancy by using stage-wise transmitter branches so that each increment is explicitly generated for its round while maintaining the same binary feedback and receiver-side stopping, thereby improving both BLER and goodput across a wide range of operating points.

\section{Conclusion} \label{sec:conclusion}
This paper advanced AE communications into an integrated PHY and link-layer framework for low latency and high reliability. By treating feedback, stopping, and retransmissions as coordinated link-layer controls, the proposed method reduces latency. Compared with baselines such as fixed-latency transmission without feedback and Chase-style repetition, the method follows a prefix-based incremental redundancy design that uses stage-wise transmitter branches to explicitly generate each redundancy increment while preserving the same binary feedback and receiver-side stopping. Results over AWGN and Rayleigh channels with multiple antenna configurations show the proposed method consistently achieves higher goodput than the baselines. Overall, lightweight link-layer control paired with a learned PHY provides a practical path to low latency operation and a foundation for broader PHY and link-layer co-design in next-generation wireless systems.

\bibliographystyle{IEEEtran}
\bibliography{references_incremental}
\end{document}